\documentclass[twocolumn,showpacs,preprintnumbers,amsmath,amssymb]{revtex4}

\usepackage{graphicx}
\usepackage{dcolumn}
\usepackage{bm}

\begin{document}

\title{Dynamics of glass phases in the two-dimensional gauge glass model}
\author{ Qing-Hu Chen$^{1,2}$, Jian-Ping Lv$^{2}$, and Huan Liu$^{2}$
}
\address{
$^{1}$ CSTCMP and Department of Physics, Zhejiang Normal University,
Jinhua 321004, P. R. China \\ $^{2}$ Department of Physics, Zhejiang
University, Hangzhou 310027, P. R. China}
\date{\today}

\begin{abstract}
Large-scale simulations have been performed on the current-driven
two-dimensional XY gauge glass model with
resistively-shunted-junction dynamics. It is observed that the
linear resistivity at low temperatures tends to zero, providing
strong evidence of glass transition at finite temperature. Dynamic
scaling analysis demonstrates that perfect collapses of
current-voltage data can be achieved with the glass transition
temperature $T_{g}=0.22$, the correlation length critical exponent
$\nu =1.8$, and the dynamic critical exponent $ z=2.0$.  A genuine
continuous depinning transition is found at zero temperature. For
creeping at low temperatures, critical exponents are evaluated and a
non-Arrhenius creep motion is observed in the glass phase.
\end{abstract}

\maketitle

\section{Introduction}

The evidences to support the existence of a  vortex glass (VG)
phase in strongly disordered type-II superconductors have been
reported in many experiments by the dynamic scaling of the
measured current-voltage data\cite{exp}. Theoretically,  the XY
gauge glass model\cite{Huse,3dgg} is often used to describe the VG
phase, although it lacks some of properties and symmetries due to
the absence of net magnetic fields\cite{Vestergren,Olsson,Lidmar}.
 Now there is general consensus that a finite-temperature VG
transition occurs in the three-dimensional gauge glass model
\cite{Kosterlitz}.

The situation is, however, much less clear in two dimensions (2D).
The  experimental quest of the VG transition in high-$T_{c}$
cuprate films\cite{Dekker,Sawa} has provided continuous excitement
and puzzles. Recently, in a positionally disordered Josephson
junction arrays with the maximal disorder strength\cite{Yun},
where the 2D gauge glass model is realized, a possible
finite-temperature glass transition has been observed
experimentally. On the theoretical side, the existence of a
finite-temperature glass transition in the 2D gauge glass model
 remains a topic of
controversy\cite{zero,dwrg,Hyman,granato,tang,li,RSJ,um,Choi,chen}.
It is predicted that, in the zero-temperature numerical
renormalization group studies of domain walls and the calculations
of stiffness exponents, there is no ordered phase at any finite
temperature in 2D\cite{dwrg,tang}. On the other hand, the
finite-temperature glass transition ($T_{g} \approx 0.2J $) has
also been spported by extensive resistively-shunted-junction (RSJ)
dynamic simulations \cite{li,RSJ,chen} and Monte Carlo simulations
\cite{Choi}.

The depinning transition at zero temperature and the creep motion at
low temperatures  have  attracted considerable attention both
analytically\cite{Nattermann,Chauve,Blatter} and
numerically\cite{Roters,luo,olsson1} in a large variety of physical
systems, such as charge density waves in solids, field-driven motion
of domain walls in ferromagnets and flux lines in type-II
superconductors. Since the non-linear dynamic response in these
systems produces a rich physical picture, there has been increasing
interest in studies of these phenomena.

In this paper, based on the RSJ dynamics, we perform large-scale
dynamic simulations on the 2D gauge glass model. Both the glass
transition temperature $T_g$ and the critical exponents are
estimated. The depinning transition at zero temperature and the
creep motion below $T_g$ are also investigated. The rest of the
paper is organized as follows. Sec.II describes the model and
dynamic method. Sec.III presents the main results, where some
discussions are also made. Finally, a short summary is given in the
last section.

\section{Model and dynamic method}

The Hamiltonian  of the 2D gauge glass model  is given
by\cite{zero}
\begin{equation}
H=-J_{0}\sum_{\langle ij\rangle }\cos (\phi _{i}-\phi
_{j}-A_{ij}), \label{Hamil}
\end{equation}
where the sum is over all nearest neighbor pairs on a 2D square
lattice, $\phi _{i}$ specifies the phase of the superconducting
order parameter on grain $i$, $J_0$ denotes the strength of
Josephson coupling between neighboring grains, and the quenched
variable $A_{ij}$ is distributed uniformly in the interval $[-\pi
,\pi )$. The present simulations are carried out with the system
size $L=128$ for all directions.

The RSJ dynamics is incorporated in simulations, which can be
described as
\begin{equation}
{\frac{\sigma \hbar }{2e}}\sum_{j}(\dot{\phi _{i}}-\dot{\phi _{j}})=-{\frac{%
\partial H}{\partial \phi _{i}}}+J_{{\rm ext},i}-\sum_{j}\eta _{ij},
\end{equation}
where $J_{{\rm ext},i}$ is the external current which vanishes
except for the boundary sites. The $\eta _{ij}$ is the thermal
noise current with zero mean and a correlator $\langle \eta
_{ij}(t)\eta _{ij}(t^{\prime })\rangle =2\sigma k_{B}T\delta
(t-t^{\prime })$. In the following, the units are taken of
$2e=J_{0}=\hbar =\sigma =k_{B}=1$.

In the present simulations, a uniform external current $I_{x}$
along the $x$ direction is fed into the system. The fluctuating
twist boundary condition \cite{chen1,kim} is applied in both
directions. The supercurrent between sites $i$ and $j$ is now
given by $\ J_{i\rightarrow j}^{(s)}=J_{0}\sin (\theta _{i}-\theta
_{j}-A_{ij}-{\bf r} _{ij}\cdot {\bf \Delta })$ with $\theta
_{i}=\phi _{i}+{\bf r}_{i}\cdot {\bf \Delta }$ and ${\bf \Delta
}=(\Delta _{x},\Delta _{y})$ the fluctuating twist variable. The
new phase angle $\theta _{i}$ is periodic in both $x$ and $y$
directions. Then, the dynamics of ${\bf \Delta }_{\alpha }$ can be
written as
\begin{equation}
\dot{\Delta}_{\alpha }={\frac{1}{L^{2}}}\sum_{<ij>_{\alpha
}}[J_{i\rightarrow j}^{(s)}+\eta _{ij}]-I_{\alpha }, \alpha =x,y.
\label{delta-dot}
\end{equation}
The voltage drop is $V=-L{\dot{\Delta}_{x}}$.

The above equations can be solved efficiently by a pseudo-spectral
algorithm \cite{chen2} due to the periodicity of the phase in all
directions. The time stepping is done using a second-order
Runge-Kutta scheme with $\Delta t=0.05$. The time-averaged voltages
are calculated over a long-time scale after reaching the steady
state. To determine the steady state, we have checked $\langle
V\rangle_{n}$ for every $(2^{n}-2^{n-1})$ time steps. We assume that
the system reaches a steady state when the fluctuation of the mean
voltage $ \left| (\langle V\rangle_{n}-\langle V\rangle _{n-1})/{
\langle V\rangle _{n}}\right|$ is less than $0.5\%$ for several
$n$'s after $n=20$. Once this criterion is satisfied, we record the
$\langle V \rangle_{n}$ as the final estimate of the voltage $V$.
The value of $n$ is typically $25$ in the present simulations. The
detailed procedure in the simulations was described in Ref.
\cite{chen2}. We have performed simulations with 10 different
realizations of disorder and observed that the results are quite
close from sample to sample, so good self-averaging effects exist in
the present large systems. This point is also supported by a recent
study in Josephson-junction arrays by Um \emph{et al.} \cite{um}.
Our results below are averaged over $10$ realizations of disorder.
For the results presented in the following figures, error bars are
smaller than or comparable with the symbol size.

\section{Simulation results and discussions}

\begin{figure}[tbp]
\centering
\includegraphics[width=6cm]{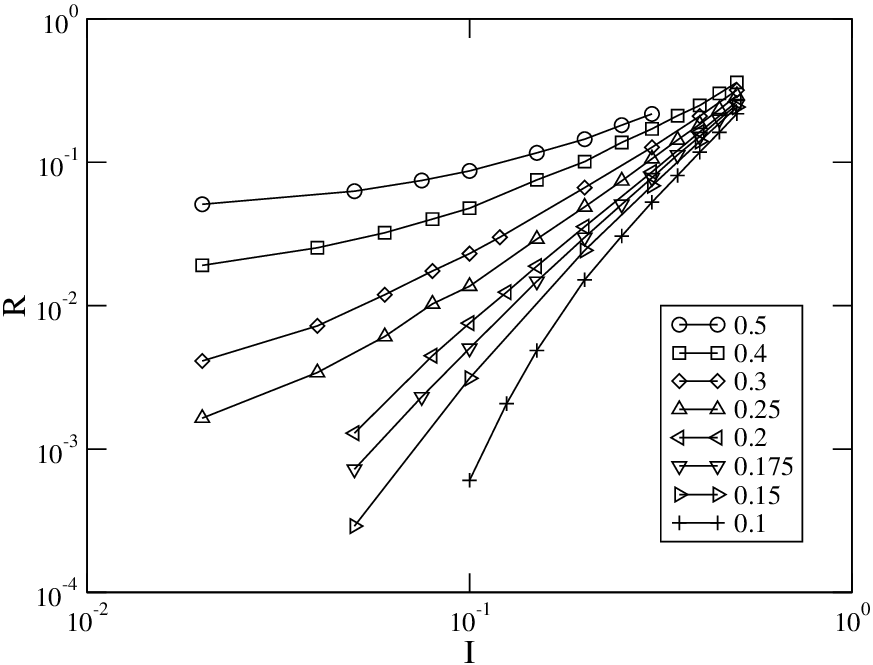}
\caption{ Log-log plots of $I-R$ curves at various temperatures.}
\label{Fig1}
\end{figure}
First, we study the possible glass phase transition. The glass
transition temperature was estimated to be $T_{g}\approx 0.20$ by
several groups\cite{RSJ,Choi,chen}. So the current-voltage
characteristics are measured at various temperatures ranging from
$0.1$ to $0.5$, which cover the previous $T_g$. At each
temperature, we try to probe the system at currents as low as
possible. Fig.\ref{Fig1} displays the resistivity $R=V/I$ as a
function of current $I$ at various temperatures in a log-log
scale. It is obvious that, at lower temperatures, $R$ tends to
zero as the current decreases, suggesting that there is a true
superconducting phase with zero linear resistivity. While at
higher temperatures, $R$ tends to a finite value, corresponding to
 Ohmic resistivity in the vortex liquid. These observations
reveal strong evidence of the existence of the low-temperature
glass phase in the 2D gauge glass model.

Assuming  that the VG transition is continuous and characterized
by the divergence of the characteristic length and time scales
$t\sim\xi^{z}$ ( $z$ is the dynamic exponent), Fisher, Fisher, and
Huse \cite {FFH} proposed the following dynamic scaling ansatz to
analyze the glass transition from a vortex liquid with Ohmic
resistivity to a superconducting glass state
\begin{equation}
TR\xi^{z+2-d}=\Psi_{\pm }(I\xi^{d-1}/T), \label{ffh}
\end{equation}
where $d$ is the dimension of the system ($d=2$ in this paper),
 $\xi\propto\mid T/T_g-1\mid^{-\nu}$  is the correlation length
which diverges at the transition and $\Psi_{\pm}(x)$ are scaling
functions for $T>T_g$ and $T<T_g$, respectively. Eq.(\ref{ffh}) is
often used to scale the measured current-voltage data in the VG
transitions in experiments.

\begin{figure}[tbp]
\centering
\includegraphics[width=6cm]{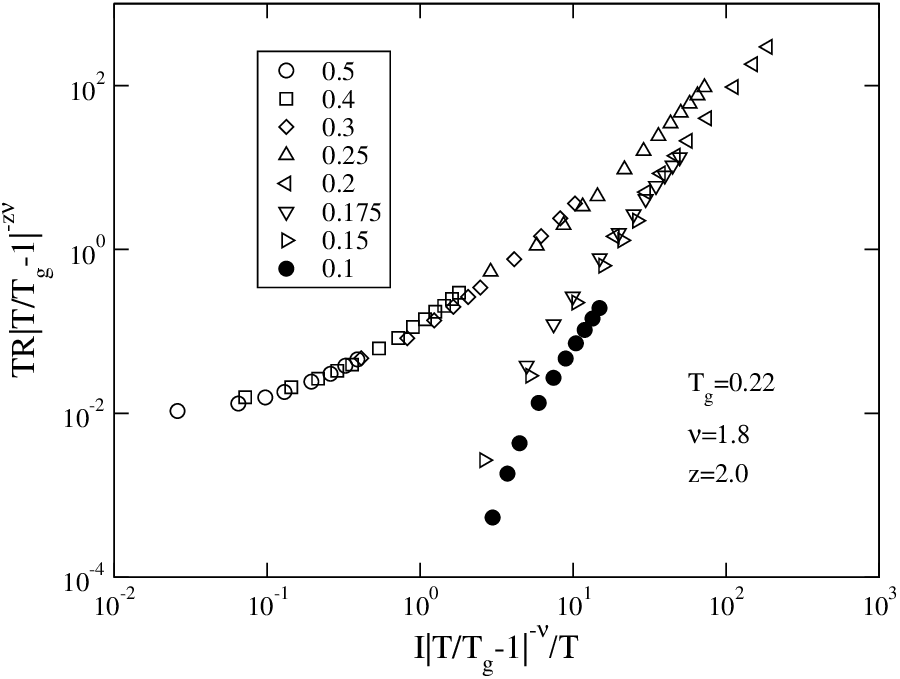}
\caption{Dynamic scaling of $I-R$ data  at various
temperatures according to Eq. (\ref{ffh}). }
 \label{Fig2}
\end{figure}

To extract the critical behavior from the numerical results of the
current-voltage characteristics, we also perform a dynamic scaling
analysis.  As shown in Fig.\ref{Fig2}, with $T_{g}=0.22\pm 0.02$ , $
z=2.0\pm 0.1$, and $\nu =1.8\pm 0.1$, an excellent collapse is
achieved according to Eq.(\ref{ffh}) except for the curve of
$T=0.1$. The errors are estimated by tuning these critical values
until the collapses become poor evidently.  The curve at $T=0.1$ is
obviously beyond the critical regime.

\begin{figure}
\centering
\includegraphics[width=6cm]{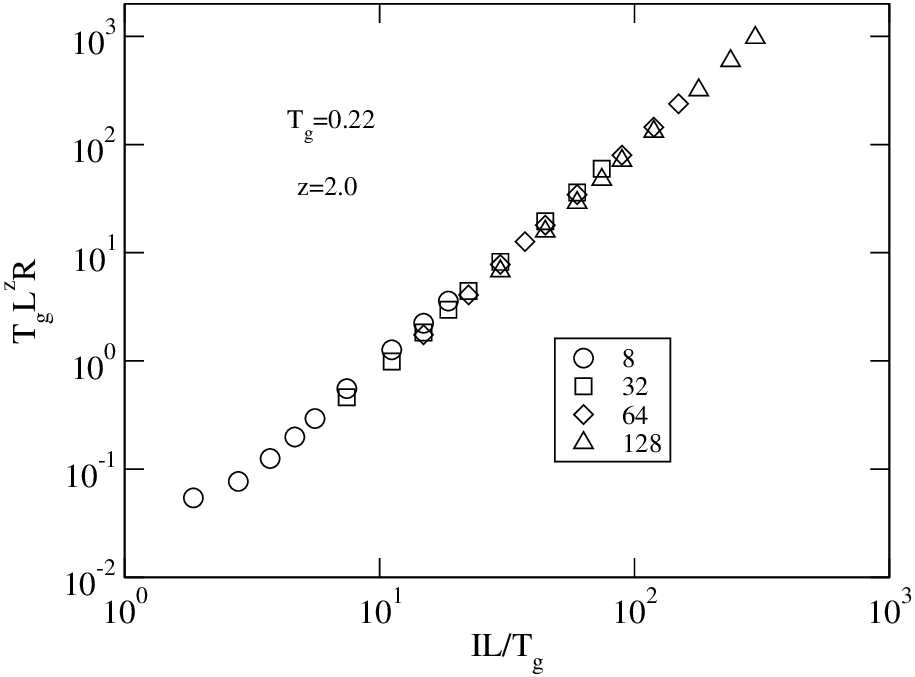}
\caption{Dynamic scaling of $I-R$ data  at
$T_{g}=0.22$ according to Eq.(\ref{ffhtg}).} \label{Fig3}
\end{figure}

The finite-size effects are particularly significant at
temperatures sufficiently close to $T_{g}$ when the correlation
length exceeds the system size. For the temperatures considered
here and the very large system size $L=128$, we believe that the
finite-size effects are negligible in the present simulations. To
confirm this point, we perform particular simulations right at
$T_{g}=0.22$ obtained above for different system sizes. At $T_g$,
the correlation length is cut off by the system size in any finite
system, so the scaling form Eq.(\ref{ffh}) becomes
\begin{equation}
T_{g}RL^{z}=\Psi (IL/T_{g}). \label{ffhtg}
\end{equation}
 A good collapse is illustrated in  Fig.\ref{Fig3} with $z=2.0$. This consistency
demonstrates that the results estimated from Fig.\ref{Fig2} are reliable.
Therefore new evidence of a finite-temperature glass transition is
provided convincingly in the 2D gauge glass model.

The obtained $T_g$ and dynamic exponent $z$ are well consistent with
those in equilibrium RSJ simulations\cite{RSJ} and Monte Carlo
simulations\cite{Choi}. The value of $\nu$ estimated here is larger
than those ( $1.1\sim 1.2$) obtained in several
simulations\cite{Choi, RSJ}, but still falls in the range of
$1.0\sim2.0$ usually observed at the VG transitions experimentally.
Within the same RSJ dynamics, the finite-size scaling  for the
linear resistivity for sample sizes $L\le 10$ gives $\nu=1.2(2)$ in
Ref.\cite{RSJ}. The discrepancy may originate from the different
scaling method and sizes used. In a previous conference paper
\cite{chen} by one of the present author and collaborators, with the
uses of a different simulation approach and a scaling form different
from Eq. (\ref{ffh}) slightly by removing temperature $T$,
$T_{g}=0.22$ , $ z=2.0$, and $\nu =1.2$ were obtained.  The present
simulation gives larger value of $\nu$.

Based on the analysis of data at temperatures above $T_g=0.22$,
previous RSJ simulations\cite{Hyman,granato} with an open boundary
condition in the 2D gauge glass model demonstrated a
zero-temperature criticality. It has been shown that the voltage
drop next to the boundary regime is particularly large and
dominates the total voltage drop across the sample at low
currents\cite{kim,simkin,Choi1}. Therefore, one should measure the
voltage drop inside the sample\cite{Choi1,chen1}. So the
conclusion based on the total voltage drop across the system with
the  open boundary condition\cite{Hyman,granato} may not be
reliable.

Interestingly, in experiments on Nb wire networks\cite{Liang},
the critical exponents $\nu =1.7\sim1.9$ were obtained for high
filling factors $f=1/2, 0.618$, and $2/5$, which are very close to
the present value. It was suggested in Ref. \cite{Yun} that the
superconducting state and transitions in the networks become
independent of $f$ in the gauge glass limit. The further work is
needed to clarify the relation between the experimental  observations
 and the present simulations.

To shed some light on the nature of this low-temperature glass
phase, we will study the depinning and creep phenomena.

\begin{figure}[tbp]
\centering
\includegraphics[width=6cm]{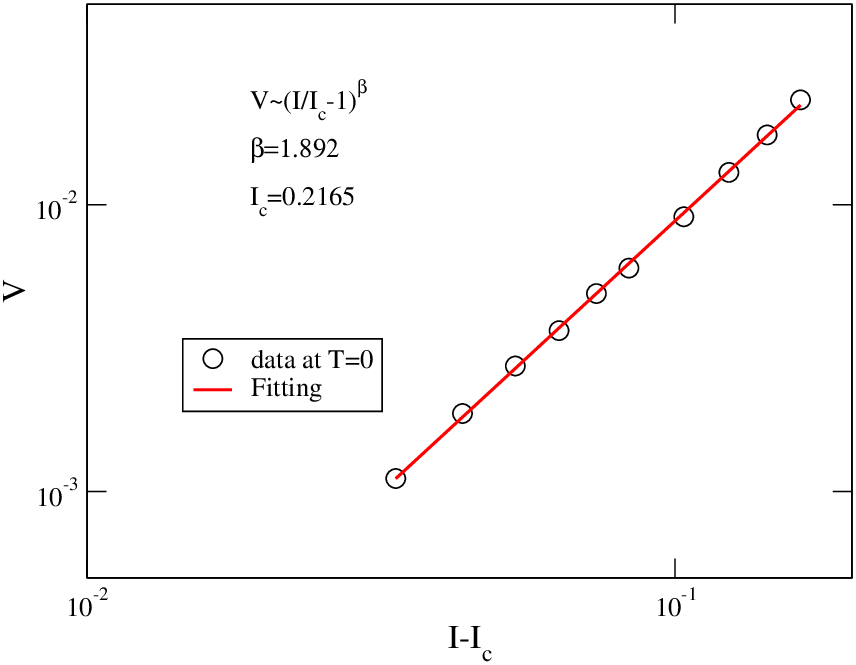}
\caption{Log-log plot of $V$ versus $(I-I_{c})$ curve at zero
temperature.} \label{Fig4}
\end{figure}

At zero temperature, we start from high currents with random initial
phase configurations.  The currents are then lowered step by step.
The steady-state phase configurations obtained at higher currents are
chosen to be the initial phase configurations of lower currents
in the next step. It becomes more and more difficult to measure the voltage
with decreasing currents. In the vicinity of the critical current, a
huge amount of computer time is consumed to get accurate results.
Fig.\ref{Fig4} presents the current-voltage characteristics at
$T=0$ in a log-log scale. We observe a continuous depinning transition
with a unique depinning current\cite{Fisher}, which can be described
as $V\propto (I-I_{c})^\beta$ with $I_{c}=0.2165\pm 0.0005$ and
$\beta=1.892\pm 0.003$. Note that the depinning exponent $\beta$ is
greater than $1$, consistent with the mean  field studies of charge
density wave models\cite{Fisher}.

At low temperatures,  the current-voltage characteristics are rounded
near the zero-temperature critical current due to thermal
fluctuations. An obvious crossover between the depinning and creep
motion can be observed around $I_{c}$ at lower accessible
temperatures. In order to address the thermal rounding of the
depinning transition, Fisher\cite{Fisher} first suggested to map
this system to ferromagnets in fields where the second-order
phase transitions occur. This mapping  was later extended to the
random-field Ising model\cite{Roters} and flux lines in type-II
superconductors\cite{luo}. If the voltage is identified as the order
parameter, the current and temperature are equivalent to the inverse
temperature and the field in ferromagnetic systems, respectively,
analogous to the second-order phase transitions, a scaling relation
among the voltage, current  and temperature in the present model
should follow the form
\begin{equation}~\label{scaling1}
V(T,I)=T^{1/\delta}S[T^{-1/\beta\delta}(1-I_{c}/I)],
\end{equation}
where $S(x)$ is a scaling function with $S(x\rightarrow 0)=$const.

\begin{figure}
\centering
\includegraphics[width=6cm]{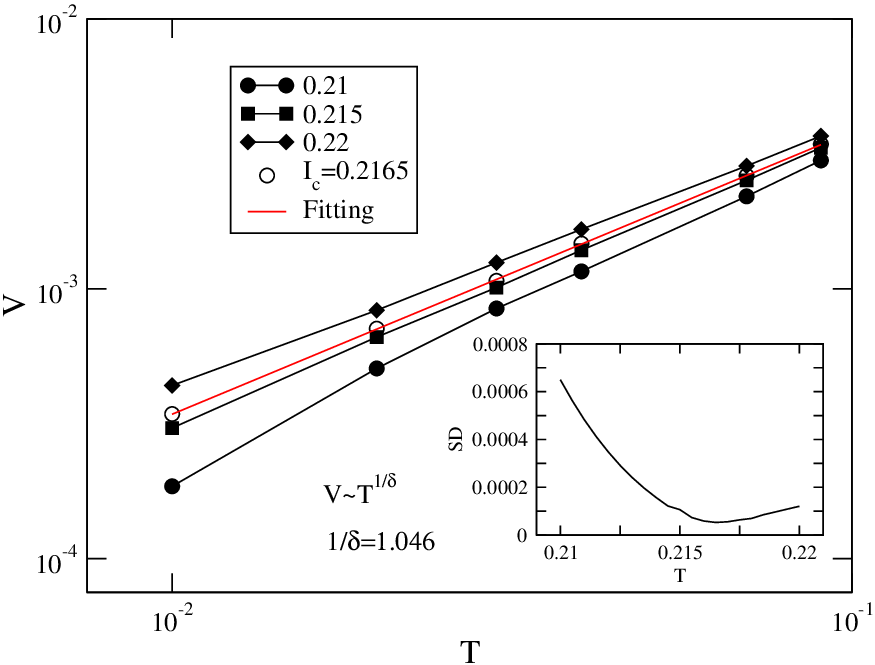}
\caption{Log-log plots of $V-T$ curves at three currents around
$I_{c}$.} \label{Fig5}
\end{figure}

It is implied in Eq.(\ref{scaling1}) that right at $I=I_{c}$ the
voltage shows a power-law behavior $V(T,I=I_{c})\propto
T^{1/\delta}$,  providing a tool to determine the critical exponent
$1/\delta$. The log-log $V-T$ curves are plotted in Fig. \ref{Fig5}
at three currents around $I_c$. We can see that the critical current is between
$0.21$ and $0.22$. The values of voltage at other currents within
$(0.21,0.22 )$ can be evaluated by quadratic interpolation. The
square deviations from the power law can be calculated. The
current at which the square deviation is minimum can be considered as
the critical current $I_{c}=0.2165\pm 0.0005$, consistent with that obtained at
zero temperature. The temperature dependence of voltage at the
critical current is also exhibited in Fig.\ref{Fig5}, yielding
$1/\delta=1.046\pm 0.002$.

\begin{figure}[tbp]
\centering
\includegraphics[width=6cm]{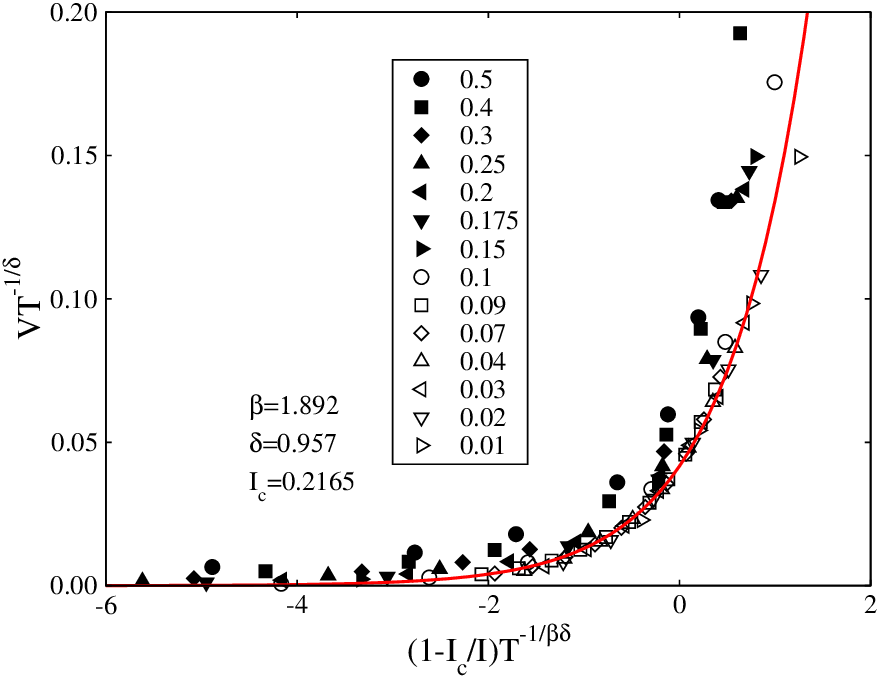}
\caption{Scaling plot  of $I-V$ data at  various
temperatures according to Eq.(\ref{scaling1}).} \label{Fig6}
\end{figure}

With the values of  $\beta$, $\delta$ and $I_{c}$ obtained above,
according to the scaling relation Eq.(\ref{scaling1}), a scaling
plot of the simulated current-voltage data in a wide range of
temperature is presented in Fig.\ref{Fig6} without any adjustable
parameter. A perfect collapse of the data for temperatures $T\le
0.10$, far below $T_g=0.22$, to a single curve for currents less
than $I_c$ is clearly shown. This collapse can be fitted well to an
exponential function $y=0.0417$exp$(1.17x)$, which is also plotted
in the Fig.\ref{Fig6} with a solid line. Note that the product of
the two exponents $\beta\delta$ describes the temperature dependence
of the creep law. Interestingly, $\beta\delta \approx 1.81$ deviates
from unity, demonstrating that the creep law is a non-Arrhenius
type. At higher temperatures, say $T>0.1$, deviations from the
scaling relation are also observed in Fig.\ref{Fig6}, which can be
attributed to strong thermal fluctuations. The non-Arrhenius type
creep phenomena only take place  at low temperatures.

\section{Summary}

We have performed large-scale dynamic simulations of the 2D gauge
glass model within the  RSJ dynamics. The strong evidence of the
low-temperature glass phase is provided in the dynamic sense. By the
dynamic scaling analysis, two perfect collapses of simulated
current-voltage data are achieved with $T_{g}=0.22\pm 0.02$ , $
z=2.0\pm 0.1$, and $\nu =1.8\pm 0.1$. The values of $T_{g}$ and $z$
are in  agreement with those in the previous equilibrium Monte Carlo
simulations. While the value of $\nu$ is larger than that  in
literature, which is, however, closer to that in experiments in
the gauge glass limit. We have also studied the depinning transition at
zero temperature and creep motion at low temperatures in detail. A
genuine continuous depinning transition is observed at zero
temperature. With the notion of scaling and the critical exponents
obtained from the simulations at zero temperature and at the critical
current, a perfect collapse of the current-voltage data at low
temperatures  is exhibited.  The value of $\beta\delta$ deviates
from unity and the scaling curve is fitted well by an exponential
function, suggesting a non-Arrhenius type creep motion in the glass
phase of the 2D gauge glass model.

It is worthy to note that in this model the current-voltage
characteristics in the whole temperature regime below $T_g$ can
almost be described in the framework of two critical phenomena. One
is the thermal rounding of the depinning transition, which is a
second-order-like phase transition. The other is the celebrated VG
transition. Further experimental and theoretical studies are needed
to clarify the relation between the present observations and
experimental findings.

\section{Acknowledgements}

We acknowledge useful discussions with X. Hu and M. B. Luo. This
work was supported by National Natural Science Foundation of China
under Grant No. 10574107 and 10774128, PNCET and PCSIRT  in
University in China, National Basic Research Program of China (Grant
No. 2006CB601003).

\end{document}